\documentclass[prd]{revtex4}%
\usepackage{amsfonts}
\usepackage{amsmath}
\usepackage{amssymb}
\usepackage{graphicx}%
\providecommand{\U}[1]{\protect\rule{.1in}{.1in}}

\begin{document}
\title[Gravitational waves in expanding universes]{Cylindrical gravitational waves in expanding universes: Models for waves from
compact sources}
\preprint{ }
\author{Robert H. Gowdy}
\affiliation{Department of Physics, Virginia Commonwealth University, Richmond, VA 23284-2000}
\author{B. Douglas Edmonds}
\affiliation{Department of Physics, Virginia Commonwealth University, Richmond, VA 23284-2000}
\author{}
\affiliation{}

\begin{abstract}
New boundary conditions are imposed on the familiar cylindrical gravitational
wave vacuum spacetimes. \ The new spacetime family represents cylindrical
waves in a flat expanding (Kasner) universe. \ Space sections are flat and
nonconical where the waves have not reached and wave amplitudes fall off more
rapidly than they do in Einstein-Rosen solutions, permitting a more regular
null inifinity.

\end{abstract}
\volumeyear{2007}
\volumenumber{75}
\issuenumber{8}
\eid{10.1103/PhysRevD.75.084011}
\received[Received 29 September, 2006]{}

\published[Published 5 April, 2007]{}

\startpage{084011}
\endpage{ }
\maketitle

This paper shows how to construct exact solutions for compact cylindrical
gravitational wave pulses spreading in an expanding universe that is spatially
flat at large cylindrical radius, where the waves have not reached. The
universal expansion is restricted to the translation-direction (or z-axis) of
cylindrical symmetry. The remaining directions show gravitational waves
spreading in a three-dimensional spacetime that can be exactly Minkowskian
before the wave pulse. The effective three dimensional spacetime has a
completely regular conformal compactification so that a radiation zone at
future null infinity is well defined without any rescaling of fields. This
simple behavior makes these solutions ideal for testing the numerical
simulation codes that are used to predict gravitational waves from compact
sources. Because they have the simplicity of Einstein-Rosen solutions but are
compatible with the boundary conditions of the real universe, these new
solutions may model localized inhomogeneities in Big Bang cosmologies and may
also provide an interesting restricted framework for testing formal ideas such
as quantum gravity.

The original Einstein-Rosen cylindrical wave solutions of Einstein's vacuum
field equations\cite{ERwaves,CylOrig} inhabit a universe that is conical
rather than flat at infinite cylindrical radius\cite{ERwavesReal} and which,
apart from the oscillations of the waves, does not evolve in time. Thus, they
are not good models for inhomogeneities in our own universe, which is thought
to be spatially flat and expanding and do not make convenient test cases for
numerical simulations of gravitational waves from compact sources. However,
the boundary conditions can be changed. Cylindrically symmetric (G$_{2}$)
exact solutions with closed space sections of $S^{3}$, $S^{1}\times S^{2}$ and
$S^{1}\times S^{1}\times S^{1}$ (three-torus) topologies are known and have
proven useful as toy models of inhomogeneous
cosmologies.\cite{gowdyvac,gowdylet,Gowdy74} The three-torus solutions, in
particular, can be regarded as standing-gravitational-wave inhomogeneous
generalizations of spatially flat Kasner
universes\cite{OscSingGowdy,GowdyPhenom,GowdyPhenom2,BergerMoncrief} and have
provided useful test cases for validating numerical simulations of Einstein's
field equations.\cite{testbeds,testbedExample} \ Here, we will change the
boundary conditions again.

We will use the notation of earlier papers\cite{gowdyvac,gowdylet,Gowdy74} on
these spacetimes, but with some changes in coordinate names to fit the new
circumstances. The isometry group coordinates are $z$ and $\varphi$ (instead
of $\sigma$ and $\delta$ used in the earlier papers), with $z$ ranging from
$-\infty$ to $+\infty$ and $\varphi$ ranging from $0$ to $2\pi$. The remaining
coordinates $t,r$ each range from $0$ to $\infty$. The isometry group consists
of translations of $z$ and of $\varphi$ modulo $2\pi$. Thus, each group orbit
is an intrinsically flat cylinder.

The advanced and retarded time coordinates, $u=t+r$ and $v=t-r,$ govern the
propagation of waves in these geometries. It is convenient to use subscripts
to denote derivatives with respect to advanced and retarded time so that, for
example, $R_{+}=\partial R/\partial u$ and $R_{-}=\partial R/\partial v$. The
spacetime metric in these coordinates is%
\begin{equation}
ds^{2}=e^{2a}\left(  dr^{2}-dt^{2}\right)  +R\left(  e^{2\psi}dz^{2}%
+e^{-2\psi}d\varphi^{2}\right)  . \label{metric}%
\end{equation}
Two of the functions in this form of the metric satisfy wave equations,%
\begin{equation}
R_{+-}=0, \label{RwaveEqn}%
\end{equation}
and%
\begin{equation}
\left(  R\psi_{+}\right)  _{-}+\left(  R\psi_{-}\right)  _{+}=0,
\label{PsiWaveEqn}%
\end{equation}
while the remaining function, $a,$ is determined by integrating the equations%
\begin{equation}
R_{+}a_{+}=R\psi_{+}^{2}+\frac{1}{2}R_{+}-\frac{1}{4}\frac{R_{+}^{2}}{R}
\label{RplusEqn}%
\end{equation}
and%
\begin{equation}
R_{-}a_{-}=R\psi_{-}^{2}+\frac{1}{2}R_{--}-\frac{1}{4}\frac{R_{-}^{2}}{R}.
\label{RminusEqn}%
\end{equation}

Because the area of a group orbit is proportional to $R$, the global character
of the solution is established by choosing a particular solution for $R.$ The
Einstein-Rosen cylindrical wave solutions correspond to the choice $R=r$. For
the solutions sometimes referred to as Gowdy T3 models, the choice is $R=t$.
For spacetimes with space sections of $S^{3}$ or $S^{1}\times S^{2}$ topology,
the appropriate choice is $R=\sin r\sin t$.\cite{gowdyvac,gowdylet,Gowdy74}
Here, we explore a \emph{new} choice:
\begin{equation}
R=rt. \label{areaR}%
\end{equation}
All of the orbits have zero area at $t=0$ and also at $r=0.$ This spacetime
expands from an initial singularity at $t=0$ and has a cylindrical symmetry
axis at $r=0$.

Separate a logarithmic solution of the wave equation from the function $\psi$%
\begin{equation}
\psi=-\frac{1}{2}\ln\left(  r/t\right)  +W \label{Wdef}%
\end{equation}
so that the metric becomes%
\begin{equation}
ds^{2}=-e^{2a}dt^{2}+e^{2a}dr^{2}+e^{-2W}r^{2}d\varphi^{2}+e^{2W}t^{2}dz^{2},
\label{metric2}%
\end{equation}
where $W$ satisfies the same wave equation as $\psi$. \ There will be a
conical singularity unless the function $a$ obeys%
\begin{equation}
\left.  \frac{\partial a}{\partial r}\right\vert _{r=0}=0,
\label{derivative constraint 1}%
\end{equation}
and the ratio of coefficients of the radius\ term $dr^{2}$ and the angle\ term
$r^{2}d\varphi^{2}$ is equal to its flat space value of $1$ to second order in
$r$. The result is a \emph{non-conicality constraint}$,$%
\begin{equation}
\left.  \left(  a+W\right)  \right\vert _{r=0}=0, \label{nonconical2 at 0}%
\end{equation}
and a derivative constraint on the function $W,$%
\begin{equation}
\left.  \frac{\partial W}{\partial r}\right\vert _{r=0}=0. \label{derivW}%
\end{equation}

The following general argument, which depends on the presence of an initial
singularity, shows that there should exist solutions that are spatially flat
at large cylindrical radius: Because $W$ satisfies a wave equation, its value
on a spacelike hypersurface $\Sigma$ near the initial singularity may be
specified arbitrarily. In particular, $W$ and its time derivatives can be
specified to have compact support $S$ on the hypersurface so that they vanish
everywhere else on $\Sigma$ and will therefore vanish throughout the
corresponding future domain of dependence $D^{+}\left(  \Sigma-S\right)
$.\cite{GerochD(S)} Wherever $W$ is zero, Einstein's equations imply that the
function $a$ satisfies a homogeneous wave equation and may then be zero
throughout this same domain of dependence. Thus, it should be possible to have
a compact wave source in a spacetime whose metric at large distance from the
source has the Flat Kasner form\cite{Kasner-sols}
\begin{equation}
ds^{2}=-dt^{2}+dr^{2}+r^{2}d\varphi^{2}+t^{2}dz^{2}, \label{FlatKasner}%
\end{equation}
which is the result of setting $a=W=0$ in (\ref{metric2}). \ (This form of the
metric is discussed further near the end of this paper.) The constant-$z$
surfaces in such a spacetime are intrinsically flat and do not show the
conical behavior that is familiar from the Einstein-Rosen cylindrical wave
solutions. In order to achieve this regularity, it must be possible for the
function $a$ to satisfy a \emph{second non-conicality constraint}; $\left.
a\right\vert _{r\rightarrow\infty}=0$.

Now construct such a solution. From (\ref{PsiWaveEqn}) and (\ref{Wdef}), the
function $W$ obeys the wave equation%
\begin{equation}
\frac{1}{t}\frac{\partial}{\partial t}t\frac{\partial W}{\partial t}-\frac
{1}{r}\frac{\partial}{\partial r}r\frac{\partial W}{\partial r}=0
\label{WwaveEqn}%
\end{equation}
for a scalar field in the Flat Kasner metric (\ref{FlatKasner}) above, with
solutions%
\begin{equation}
W=\int dk\left[  A\left(  k\right)  J_{0}\left(  kt\right)  +C\left(
k\right)  N_{0}\left(  kt\right)  \right]  J_{o}\left(  kr\right)  ,
\label{waveEqn}%
\end{equation}
where $J_{0}$ and $N_{0}$ are Bessel functions of the first and second kind.
The remaining metric function, $a$, is determined by integrating equations
(\ref{RplusEqn}) and (\ref{RminusEqn}), which now become%
\begin{equation}
ua_{+}=\frac{1}{2}\left(  u^{2}-v^{2}\right)  W_{+}^{2}-vW_{+}
\label{aPlusEqn}%
\end{equation}
and%
\begin{equation}
-va_{-}=\frac{1}{2}\left(  u^{2}-v^{2}\right)  W_{-}^{2}+uW_{-}.
\label{aMinusEqn}%
\end{equation}
This last equation implies a matching constraint at $v=0$ or $r=t$ of the same
sort that is encountered in the $S^{3}$ and $S^{1}\times S^{2}$
solutions.\cite{gowdyvac,gowdylet,Gowdy74}
\begin{equation}
\left.  rW_{-}\right\vert _{r=t}=0;-1. \label{MatchingConstraint}%
\end{equation}
Note that the quadratic nature of (\ref{aMinusEqn}) yields two solutions and
thus two possible forms of the matching constraint. This constraint is
preserved in time when the field equations are satisfied. Impose it on the
solution given by (\ref{waveEqn}) and obtain an expression that involves only
the Wronskian of the Bessel functions. Abel's formula for the
Wronskian\cite{TenenbaumODE,AbelsFormula} then yields a constraint on the
coefficients $C\left(  k\right)  :$%
\begin{equation}
\int dk\frac{C\left(  k\right)  }{k}=0;\pi. \label{Abel}%
\end{equation}
Here, we can see that one possible form of the matching constraint permits
$C\left(  k\right)  =0$ while the other requires $C\left(  k\right)  \neq0$.
When either form of this constraint is satisfied, the function $a$ can be
expressed as the integral%
\begin{equation}
a\left(  x,t\right)  +W\left(  x,t\right)  =\int_{0}^{x}\frac{t^{2}Z}%
{t^{2}-r^{2}}rdr, \label{aPlusW}%
\end{equation}
where%
\begin{equation}
Z=\left(  W_{r}^{2}+W_{t}^{2}\right)  -2\frac{r}{t}W_{r}W_{t}-2\frac{r}{t^{2}%
}W_{r}+\frac{2}{t}W_{t}. \label{Zdef}%
\end{equation}

In the case of interest, with $W\left(  x,t\right)  =0$ for $x\gg t,$ the
requirement of a non-conical metric in the part of spacetime not reached by
gravitational waves means that there is an integral constraint on the initial
data for the wave:%
\begin{equation}
\int_{0}^{\infty}\frac{t^{2}Z}{t^{2}-r^{2}}rdr=0. \label{IntegralConstraint}%
\end{equation}
This constraint is quite similar in form to the integral constraints that are
encountered in the spatially closed G$_{2}$ solutions. In those cases, the
constraint is needed to have a metric that is non-conical at both cylindrical
symmetry axes or, for the three-torus solutions, a metric that is periodic in
space. Here, one of the `axes' is at infinity but the situation is otherwise
the same.\cite{gowdyvac,gowdylet,Gowdy74}

Notice that for $t>r$ the function $Z$ is dominated by the positive definite
expression $\left(  W_{r}^{2}+W_{t}^{2}\right)  $ and can be thought of as an
energy density. In that region, the solution is just a standard Einstein-Rosen
cylindrical wave in distorted coordinates. At large values of $t,$ the
coordinates go over to the canonical Einstein-Rosen coordinates and the
integrand becomes what has sometimes been called the \textquotedblleft
C-energy\textquotedblright.\cite{Cenergy} The positive definite nature of the
C-energy is the reason that standard Einstein-Rosen cylindrical wave solutions
must be conical at large cylindrical radius.

In the region where $r>t$, the integrand reverses sign and the dominant term
of $Z$ becomes $W_{r}W_{t}$, which has the form of a scalar field energy-flux
and can have either sign. In this region, our solution is just a distorted
form of the cosmological Einstein-Rosen solution (sometimes identified as a
Gowdy $T^{2}$ or $T^{3}$ toroidal spacetime) that expands from an initial
singularity. In the canonical coordinates for this type of solution, the
analog of our integral for the function $a$ contains only an energy-flux type
of term. The form of the integral constraint makes it clear that there must
always be some non-zero wave in the $r>t$ region. A wave with compact initial
data near $r=t=0$ would straddle the $r=t$ surface with parts in both regions.
A wave pulse with compact initial data near $r=0$ at a later time would need a
precursor pulse with $r>t$ to satisfy the constraint.

A well known disadvantage of Einstein-Rosen cylindrical waves is that their
amplitudes fall off as $r^{-1/2}$ at large cylindrical radius $r$, which
prevents them from being regular at future null infinity. The difficulty can
be managed by defining rescaled wave amplitudes,\cite{CauchyChar1} but that
special treatment complicates the use of cylindrical waves as test cases for
formalisms and numerical methods designed to treat three dimensional waves.
The solutions discussed here acquire an additional factor of $t^{-1/2}$ from
their Bessel function time dependence so that wave packets with support near
the $r=t$ surface fall off as $1/r$.

A simple way to examine future null infinity is to apply the compactified
coordinates, $\tau$ and $\theta$ defined in the usual way by $u=\tan\frac
{1}{2}\left(  \tau+\theta\right)  $ and $v=\tan\frac{1}{2}\left(  \tau
-\theta\right)  $ to the three dimensional spacetime perpendicular to the
symmetry axis.\cite{relGrpsTopPenrose,penrosediag} The metric $^{3}ds^{2}$ of
this spacetime is brought into the form $^{3}ds^{2}=\Omega^{-2}d\tilde{s}^{2}$
where $\Omega=2\cos\frac{1}{2}\left(  \tau+\theta\right)  \cos\frac{1}%
{2}\left(  \tau-\theta\right)  $ and%
\begin{equation}
d\tilde{s}^{2}=e^{2a}\left(  d\theta^{2}-d\tau^{2}\right)  +\sin^{2}\theta
e^{-2W}d\varphi^{2}, \label{ConformalMetric}%
\end{equation}
so that the spacetime is embedded as the open set defined by $0<\tau<\pi$,
$0<\tau+\theta<\pi$ in a cylindrical $S^{2}\times S^{1}$ spacetime. Future
null infinity is then the set of points described by $\tau+\theta=\pi$. For
Einstein-Rosen waves one finds that $W$ has the form $W\rightarrow\Omega
^{1/2}f$ where $f$ is a smooth function, so that derivatives of $W$ will
contain $\Omega^{-1/2}$ and will diverge at future null infinity. Thus,
Einstein-Rosen waves do not admit a regular compactification of the spacetime
perpendicular to the symmetry axis. For the solutions discussed here, $W$ has
the form $W\rightarrow\Omega f$ and is clearly regular at future null
infinity. Because the derivatives of the function $a$ are related to the
derivatives of $W$ by (\ref{aPlusEqn}) and (\ref{aMinusEqn}) that function is
also regular at future null infinity. The solutions presented here have a
regular compactification with no need for rescaling.

A more complete analysis of null infinity in Einstein-Rosen spacetimes
indicates that the full four-dimensional spacetime is somewhat better behaved
than the three-dimensional section perpendicular to the symmetry
axis.\cite{ERasymp1,ERasymp2} It would be interesting to see a similar
analysis of the solutions presented here. It is worth noting that null
infinity for the toroidal model spacetimes has also been
analyzed.\cite{toroidalNull,moreToroidal}

The Flat Kasner metric given by (\ref{FlatKasner}) serves as the background
spacetime for these gravitational waves. Here, we choose to regard it as an
axially symmetric Bianchi type I homogeneous spacetime expanding from a
singularity at $t=0$. However, it can be characterized in several other ways
that may be significant for a complete analysis of null infinity in these
solutions. It has zero spacetime curvature and is similar to the Milne
cosmology.\cite{MilneCosmology} It is the sector of Minkowski spactime that
corresponds to the future of the $t=z=0$ two-surface and is presented in
Rindler hyperbolic coordinates. \ The symmetry in our $z$ direction
corresponds to Lorentz boosts along one Minkowski space axis. If our $z$
coordinate is made to be periodic, the result is Misner
space\cite{MisnerSpace}. With additional periodicities, the result is the
background geometry of the toroidal Gowdy T3 spacetimes.The fact that the
Milne and Misner spaces have long been favorite toy models of Big Bang
universes adds to the possible usefulness of our solutions.

The solution for the wave function $W$ given by (\ref{waveEqn}) may not always
be the most convenient format to use. For example, one might wish to give
Cauchy data at an initial time instead of amplitude functions. That problem
has been solved by Woodhouse for the conventional Einstein-Rosen waves and a
similar analysis may be possible in this, somewhat different
case.\cite{WoodhouseERTwistor} Again, one might wish for a less general set of
solutions that are given in terms of elementary functions so that one can
avoid the need to perform integrals. That problem was solved for the
conventional Einstein-Rosen waves by the Weber-Wheeler pulse solution and a
similar specialization should be possible here as well.\cite{ERwavesReal}

The solutions presented here have just one polarization of gravitational waves
and no matter fields. It is well known that a second polarization can be added
in the form of an off-diagonal term in the metric\cite{gowdyvac} and massless
matter fields such as an electromagnetic wave\cite{EMGowdy} and a massless
scalar or axion field\cite{ScalarGowdy,KunzeStringCosm} can also be added
without changing the basic structure of these spacetimes. The effect of the
added degrees of freedom is to replace the linear wave equation
(\ref{PsiWaveEqn}) or (\ref{WwaveEqn}) by a system of nonlinearly coupled wave
equations. These nonlinear generalizations have provided a simple setting for
numerical explorations of nonlinear gravitational
interactions.\cite{BergerMoncrief,GowdyPhenom,GowdyPhenom2} Analytic
procedures for generating solutions to this nonlinear system, as it appears in
canonical Einstein-Rosen coordinates, have also been
developed.\cite{AlxseevSolGenER}

Because the Einstein-Rosen solutions and their nonlinear generalizations are
the simplest examples of gravitational waves that radiate outwards, they have
been used to test both formal ideas and numerical procedures in a simple
setting. There is, for example, a long history of using them as a framework
for testing ideas about quantum
gravity.\cite{KucharQuantER,quantERscalar,KunzeStringCosm} They have, as
another example, been used to develop Cauchy characteristic matching
techniques for extracting gravitational radiation from numerical simulations
of astrophysical sources.\cite{CauchyChar1,CauchyChar2} Another technique to
accomplish the same purpose, the use of hyperboloidal slicing, is still being
tested on scalar waves in fixed background spacetimes but will need more
complex tests shortly.\cite{CauchyCharSphSym} The solutions presented here are
slightly more complex than the Einstein Rosen solutions. However it may be
argued that they fit the boundary conditions of our own universe while the
Einstein Rosen solutions do not. Thus, these solutions, and their nonlinear
generalizations to multiple polarizations and massless fields are a logical
step up from the Einstein Rosen solutions as test vehicles for new techniques.

\textit{Note added in proof. -- }The solutions presented here belong to the
general class of boost-symmetric spacetimes discussed by Bi\v{c}ak and
Schmidt, Phys. Rev. D \textbf{40}, 1827 (1989).

\bibliographystyle{apsrev}
\bibliography{acompat,gowdy}

\end{document}